\documentclass[aps,prapplied,twocolumn,superscriptaddress]{revtex4-2}
\usepackage{float,graphicx,multirow, amsmath,color,dsfont}
\usepackage{amsmath,bm}
\usepackage{svg}
\usepackage[colorlinks]{hyperref}
\usepackage[noabbrev]{cleveref}
\usepackage{enumerate}
\usepackage{amssymb,mathrsfs,dsfont}
\usepackage{amssymb,mathbbol}
\usepackage{amsfonts}
\usepackage[version=4]{mhchem}
\usepackage{mathrsfs}
\usepackage{cancel}
\usepackage{float}
\usepackage{siunitx}
\usepackage{physunits}
\DeclareSIUnit\au{a.u.} 
\usepackage{txfonts}
\usepackage{physics}
\usepackage{xcolor,hyperref}
\definecolor{darkblue}{rgb}{0,0.0.1,0.3}
\definecolor{darkred}{rgb}{0.6,0.1,0}
\hypersetup{colorlinks,breaklinks,
	linkcolor=darkred,urlcolor=darkblue,
	anchorcolor=darkred,citecolor=
	darkred,pdfauthor=MZI, pdftitle= MZI}
\usepackage{tikz}
\usepackage{mathbbol}
\usepackage{float}
\usepackage{array}
\usepackage{accents}

\usepackage{braket}

\usepackage[normalem]{ulem} 

\DeclareSIUnit{\sqrthz}{\ensuremath{\sqrt{\text{\hertz}}}}

\begin{document}	

\title{Quantum Imaging of Nonlinear Magnetization Reversal and Dissipative Domain Dynamics}
\author{Manali Verma}
\thanks{These authors contributed equally to this work.}
\affiliation{Department of Electrical Engineering, Indian Institute of Technology Bombay, Powai, Mumbai 400076,
Maharashtra, India}
\author{Dasika Shishir}
\thanks{These authors contributed equally to this work.}
\affiliation{Qmet Tech Foundation, IIT Bombay, Powai, Mumbai 400076, Maharashtra, India}
\author{Prajna Shubhranshu Mahakur}
\affiliation{Department of Electrical Engineering, Indian Institute of Technology Bombay, Powai, Mumbai 400076,
Maharashtra, India}
\author{P.S Anil Kumar}
\affiliation{Department of Physics, Indian Institute of Science, Bangalore 560012, India}
\author{Kasturi Saha}
\affiliation{Department of Electrical Engineering, Indian Institute of Technology Bombay, Powai, Mumbai 400076,
Maharashtra, India}
\affiliation{Qmet Tech Foundation, IIT Bombay, Powai, Mumbai 400076, Maharashtra, India}


\def\thefootnote{*}\footnotetext{These authors contributed equally to this work.}\def\thefootnote{\arabic{footnote}}

\begin{abstract}
To investigate the rich nonlinear dynamics of magnetic materials, we introduce a stroboscopic quantum diamond microscopy technique to probe the dynamic AC response of a Pt/Co/Pt thin film using nitrogen-vacancy centres in diamond. By overcoming the limited dynamic range of conventional optically detected magnetic resonance based imaging, our approach enables the observation of magnetization reversal under various large-amplitude sinusoidal and square wave excitations. The observed phase delay between the applied field and the sample response reveals information about the energy dissipated within the material. The temporal evolution reveals magnetization reversal through domain nucleation and growth also demonstrating isolated bubble domains associated with the defects. Furthermore, the detection of higher order harmonics provides access to the nonlinear AC susceptibility. These results establish stroboscopic quantum diamond microscope as a promising tool for spatially and temporally resolving nonlinear magnetization dynamics, with potential applications extending beyond ferromagnetic systems to complex magnetic crystals, magnetic phase transitions and superconductor vortex dynamics.
\end{abstract}
\maketitle

\section{Introduction}
\begin{figure*}[t]
    \centering
    \includegraphics[width=\linewidth]{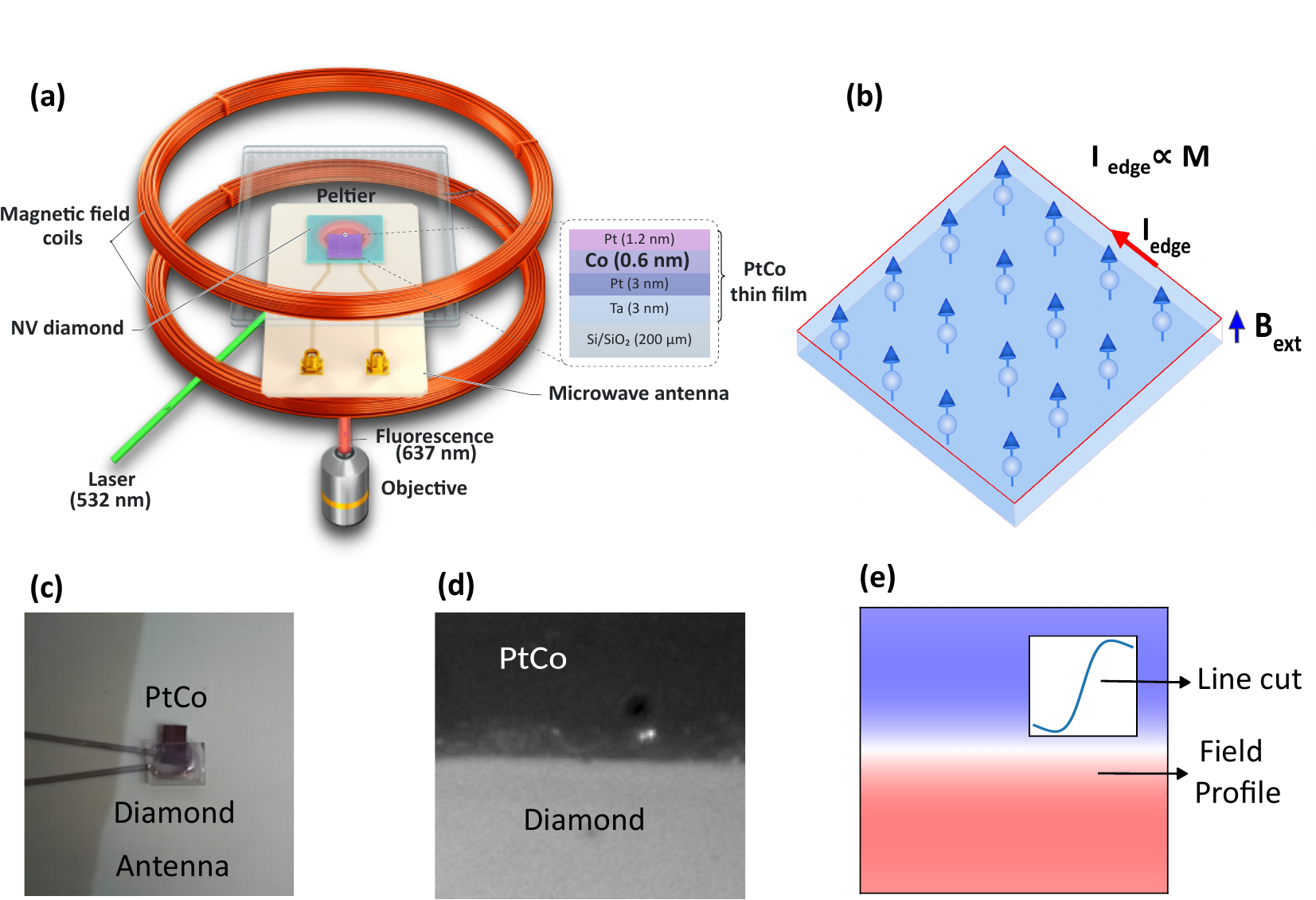}
    \caption{(a). Experimental Platform: A thin film of Pt/Co/Pt is placed in close proximity with diamond with a thin layer of \ce{NV-} centers. A  pair of current carrying coils are used to provide necessary bias magnetic field. (b). A perpendicularly magnetized film with magnetization $\vb{M}$ is  equivalent to an effective current flowing along the film edges. (c). Pt/Co/Pt mounted on the microwave antenna with the diamond sensor (d). Microscopic image showing the edge of the Pt/Co/Pt  thin film.  (e). Simulated magnetic field from the edge of the sample with the inset showing a line cut along $y$-axis.}
    \label{fig:overview}
\end{figure*} 
Visualizing magnetic dynamics play a vital role in understanding the fundamental
physics of magnetic materials\,\cite{2pff-xy6n,Sharma2024} and developing relevant
industrial applications like random access memories\,\cite{https://doi.org/10.1002/pssr.201105420}, sensors\,\cite{s22030961} and neuromorphic memories\,\cite{Marrows2024} 
around them. Among the various phenomena exhibited by ferromagnetic materials, magnetization reversal is one of the most crucial process.  For instance, 
information in magnetic data storage is encoded as binary bits (1s and 0s) represented by the direction of local magnetization. The speed, storage density and reliability of writing data depends directly on how fast and efficiently a magnetic domain can be reversed. In very small 
magnetic systems, the magnetization reversal usually takes place via coherent
rotation. However, in larger engineering materials, the reversal happens 
through domain wall motion which is a gradual process. Upon reversal of external magnetic fields, small magnetic domains get reversed and nucleate at local defects, grain boundaries, or surface roughness sites, followed by the displacement of domain walls\,\cite{cullity2011introduction}. This process is limited in time by the finite domain
wall velocity associated with a relaxation time constant, $\tau_C$. 
Visualizing and characterizing the reversal process as well as the time 
constant associated with it is hence of utmost importance for understanding the underlying magnetic mechanisms and optimizing the performance of magnetic devices.

Among the many tools available for studying microscale dynamics of magnetic 
materials, nitrogen-vacancy (\ce{NV-}) centers offer high spatial resolution, temporal resolution
and quantitative magnetic field imaging while avoiding cryogenic temperatures and vacuum 
environments\,\cite{simpson2016magneto,dasika2023mapping}.  
A substantial fraction of magnetic field studies based on \ce{NV-} have 
focused on static or direct current (DC) 
magnetometry \, \cite{sun2021magnetic,10.1063/5.0091931,casola2018probing,simpson2016magneto,kitagawa2024wide,mathesa2026investigation}. Apart from this, recent advances in coherent spin control and continuous wave double resonance spectroscopy method have extended \ce{NV-} based sensing to linear alternating current (AC) magnetometry\,\cite{dasika2023mapping,10.1063/5.0091931,casola2018probing, dasika2025quantum,zhang2021ac,kitagawa2025imaging, saijo2018ac} 
enabling the spatially resolved investigation of dynamic magnetic properties. By synchronizing the evolution of \ce{NV-} spins with an applied AC excitation, quantum diamond microscopy can quantify linear AC magnetic susceptibility \cite{zhang2021ac,dasika2023mapping,dasika2025quantum}. More recently, scanning nitrogen-vacancy center magnetometry have also been used to map nonlinear spin-wave frequency multiplication in Permalloy\,\cite{daddario2026imagingharmonicgenerationmagnons}. Such measurements provide access to magnetic dynamics that may remain obscured in conventional DC measurements. Nevertheless, characterizing the nonlinear AC response of magnetic materials, particularly through higher order susceptibility, magnetization reversal and their out of phase component remains an important avenue for exploring their dynamical behavior. In the weak field limit, the response is almost linear, and the undisturbed ground state of the magnetic system can be probed without any significant changes induced in the magnetic structure. However, exploring the nonlinear regime requires the application of a stronger external magnetic field, under which the magnetization response deviates from its linear dependence on the field strength. This nonlinearity manifests itself through the emergence of higher-order harmonics in the AC susceptibility response.

In this work, we present a novel approach for measuring and visualizing the dynamics of Pt/Co/Pt thin film, using a custom built \ce{NV-} based quantum diamond microscope (QDM) setup. The nonlinear dynamics of ferromagnetic materials under time dependent magnetic excitations encode critical information about magnetization reversal, domain evolution and higher order nonlinear response. We employ a widefield stroboscopic method to detect AC magnetic fields emanating from the thin film. These measurements pose significant challenges with conventional methodologies due to the limited dynamic range of conventional optically detected magnetic resonance (ODMR)-based magnetic imaging techniques. We demonstrate the dynamic effects seen under full hysteresis sweep, which includes phase delay in the response of magnetization of the thin film, and the higher order non linear AC dynamics at various temperatures with different driving frequencies. By repeatedly driving the magnetic thin film to positive and negative saturation, we visualize the domain wall motion and derived the associated time constant. Our method enables imaging magnetization dynamics on the scales of  \SI{}{\milli \second} with broad dynamic range and 
high spatial resolution. The resulting dynamics are further analyzed in the Fourier domain, allowing us to resolve the real and imaginary components of the magnetization response.

\section{Setup and Implementation}

The experiments are performed using a custom-built side-illumination widefield QDM setup, as shown in Fig.\,\ref{fig:overview}(a). The measurements are carried out over a field of view of $\SI{70}{\micro \meter} \cross \SI{70}{\micro \meter}$ with a per pixel sensitivity of $\eta $ = \SI[power-half-as-sqrt]{3}{\micro \tesla \per \sqrt{\hertz}}. The magnetic thin film used for measurements consist of a stack of 
Ta(\SI{3}{\nm})/Pt(\SI{3}{\nm})/Co(\SI{0.6}{\nm})/Pt(\SI{1.2}{\nm}) deposited on a \ce{Si} 
wafer using DC magnetron sputtering\,\cite{parakkat2016tailoring,constant1930magnetic}, resulting in a structure exhibiting perpendicular magnetic anisotropy (PMA). The magnetic field from the sample is sensed using a $99.99\%$ isotopically purified diamond from Element Six Co. measuring \qtyproduct{4 x 4 x 0.25}{\mm}. The diamond is enriched with \ce{^{12}C} and implanted with \ce{^{15}N}, forming a \ce{NV-} layer approximately \SI{1}{\um} thick from the diamond surface with the \ce{NV-} centers orientated along the [111] crystallographic axis.

For optical excitation of \ce{NV-} centers, a \SI{532}{\nm} optically pumped solid state laser (GEM 532) is used which is further modulated by an acousto optic modulator (AOM) (Isomet M1133-aQ80L-1.5) and directed towards the sample using a periscope setup. A Nikon $10\times$ objective is used for collecting the photoluminescence (PL), which is further imaged by the camera (Basler ace acA720-520 \SI{}{\micro \meter}). The sample is positioned and scanned precisely using an XYZ stage (Thorlabs PT3/M). Microwave control of the \ce{NV-} spins are achieved using microwave pulses from the Windfreak SynthNV signal generator which is further amplified with a high power amplifier (Mini-Circuits ZHL-16W-43S+) and delivered to the sample through a custom made RF antenna. The homogeneity of the microwave field from the antenna is 95\%.

For measurement, the thin film is placed on a Peltier heating element whose temperature can be controlled from room temperature (RT) to \SI{100}{\degreeCelsius} using a custom built PCB incorporated with Raspberry Pico microcontroller. The effect of laser heating on the sample is also measured to be \SI{0.58}{\degreeCelsius}/ \SI{100}{\milli \watt} (See Supplementary Information, Fig. S1). Two parallel current carrying coils, perpendicular to the sensor are used for applying the bias magnetic fields, generating \SI{0.35}{\milli \tesla}  for every \SI{1}{\ampere} of current applied. As the thin film has strong PMA, only perpendicular fields along $z$-axis are applied throughout the experiment described in this work. The current to the coils are applied through a custom made bipolar voltage controlled current source. The control voltage to the current source is synchronized with the applied microwave pulse using the Pulse streamer 8/2 (Swabian Instruments) for applying various sinusoidal and square waveforms.

For fields applied only along $z$-axis, the magnitude of projection of magnetic field along all the four \ce{NV-} 
orientations are equal, giving only two dips $\nu_\pm$ in the ODMR spectrum,\,\cite{Dubey2025}
\begin{equation}
\nu_\pm = D \pm \frac{\gamma_e B_{z,a+s}}{\sqrt{3}},
\label{eq:perp_field}
\end{equation}
where, $D$ is the temperature dependent zero field splitting resonance (\SI{2870}{\mega \hertz}), 
$\gamma_e$ is the electron gyromagnetic ratio (\SI{28}{\mega\hertz\per \milli\tesla}) and 
$B_{z,a+s}$ is the magnetic field  from the sample and the bias. In this magnetic field 
configuration, the \ce{NV-} centers are only sensitive to perpendicular magnetic fields. 
Fig.\,\ref{fig:overview}(b) shows a magnetic thin film with perpendicular magnetization $M$ which 
is ideally equivalent to a current carrying slab, carrying a current $I$ only along its edge. The 
equivalence
between the current $I$ and the magnetization $M$ is given by\,\cite{sadiku2018elements}
\begin{equation}
M = \frac{I}{t}
\label{eq:current_magnetization},
\end{equation}
where, $t$ is the thickness of the thin film. The diamond sensor is mounted on the magnetic thin film such that the edges of the thin film can be easily captured in the field of view as shown in Fig.\,\ref{fig:overview}(c), and (d). The resultant magnetic field from such a configuration is depicted in
Fig.\,\ref{fig:overview}(e). The inset in Fig.\,\ref{fig:overview}(e) shows the line-cut along the $x$-axis.
By fitting the line-cut to 
\begin{equation}
B_{z,s}(x) = \frac{\mu_\circ I}{2\pi} \frac{x}{x^2 + z_\circ^2},
\label{eq:B}
\end{equation}
the equivalent edge current can be extracted (I = \SI{240}{\micro \ampere}). Here, $z_\circ$ is the stand-off distance (\SI{15}{\micro \meter}) between the \ce{NV-} layer and the thin film and $\mu_\circ$ is the permeability of free space. Finally by combining Eq.\,\eqref{eq:current_magnetization}, and
Eq.\,\eqref{eq:B}, the magnetization value $M$ can be determined. 

\section{Result and Discussion}
\subsection{Static}
\begin{figure}
    \centering
    \includegraphics[width=\linewidth]{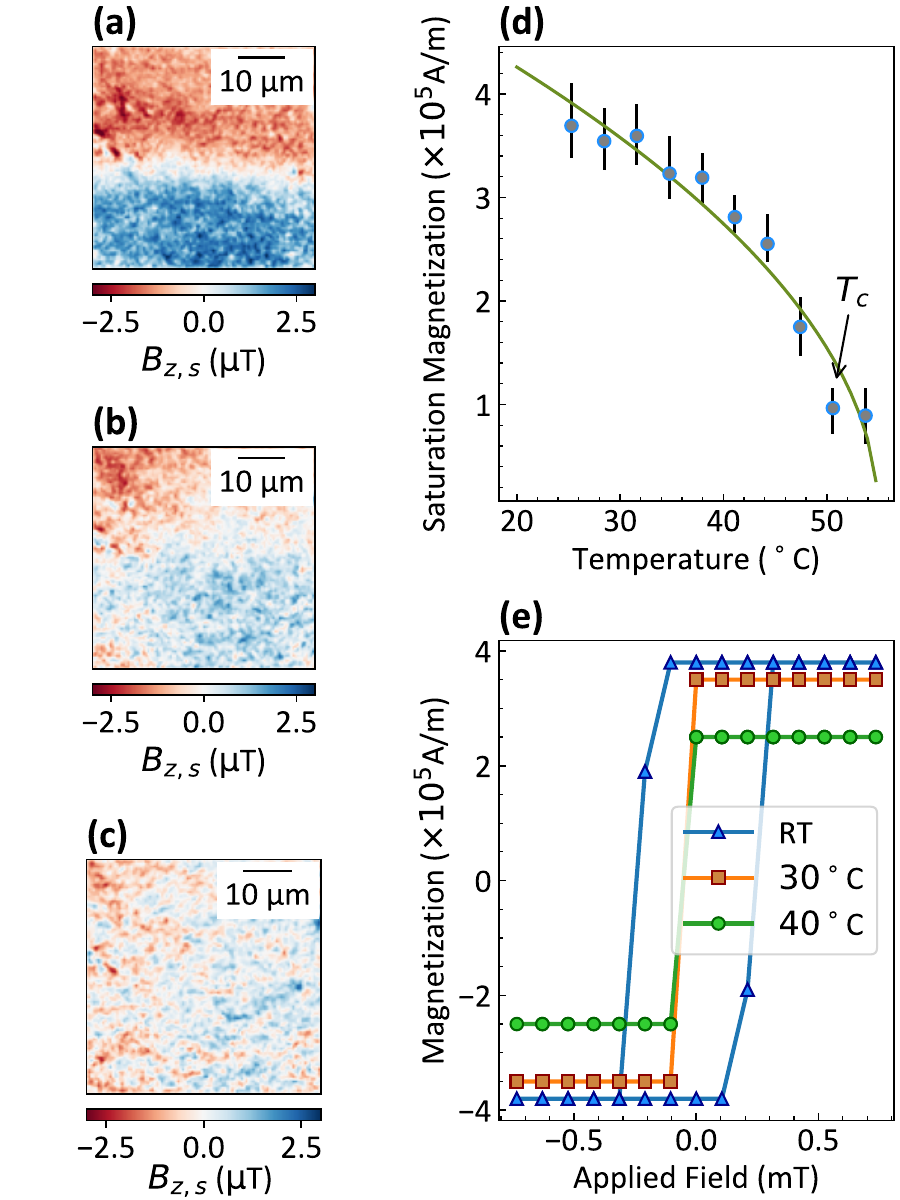}
    \caption{Static magnetic field data from the thin film at (a). \SI{24}{\celsius} (b). \SI{45}{\celsius} and (c). \SI{51}{\celsius} when \SI{0.7}{\milli\tesla} bias magnetic field is applied. (d). Saturation magnetization at various temperatures (e). MH curve at various temperatures.}
    \label{fig:dc}
\end{figure}

To investigate the magnetic properties of the thin film\,\cite{pavlov2001linear}, we first characterize its static magnetic response as a function of the applied bias magnetic field and thin film temperature. All measurements are performed at a constant laser power of \(\SI{500}{\milli\watt}\) at the source, corresponding to \(\SI{350}{\milli\watt}\) incident on the thin film, with the bias magnetic field applied perpendicular to the film plane. Fig.\,\ref{fig:dc}(a)-(c) present widefield magnetic images acquired at different temperatures, \SI{24}{\degreeCelsius}, \SI{45}{\degreeCelsius} and \SI{51}{\degreeCelsius} under a constant bias field of \(\SI{0.7}{\milli\tesla}\), revealing that the magnetization progressively diminishes with increasing temperature and finally vanishes around \SI{51}{\celsius} 
indicating that it is the Curie temperature of the thin film, above which long range ferromagnetic order is lost. To quantitatively determine the Curie temperature, we plot the variation of saturation magnetization $M$ with various temperatures $T$ under a bias magnetic field 
of \SI{0.7}{\milli \tesla} in 
Fig.\,\ref{fig:dc}(d). The MT curve is fit to the equation
\begin{equation}
M(T) = M(0) \qty(1 - \qty(\frac{T}{T_c})^{\gamma}),
\label{eq:bloch}
\end{equation}
where, $M(0)$ is the magnetization at \SI{0}{\kelvin} and $T_C$ is the Curie
temperature \cite{van1993controlling}. The fit yields a Curie temperature of \SI{50}{\degreeCelsius} and $\gamma = 0.5$ which is the critical exponent, depending on the universality class of the material. Fig.\,\ref{fig:dc}(e), shows the five quadrant hysteresis (MH) curve at various 
temperatures \cite{swallow1975properties}. At RT (\SI{24}{\degreeCelsius}), the thin film shows a binary 
hysteresis, where the magnetization is either in fully magnetized up 
state or fully magnetized down state. It requires a coercive field of approximately \SI{0.2}{\milli \tesla} to 
flip the state of magnetization with hysteresis width spanning around 0.5 mT. At higher temperatures, the thin film shows nearly no coercivity, such that even a slight magnetic field is sufficient to flip the state of magnetization towards the bias magnetic field. The saturation magnetization of the thin film at RT, \SI{30}{\degreeCelsius} and \SI{40}{\degreeCelsius} is \SI{3.7(3)e5}{\ampere\per \meter}, \SI{3.5(2)e5}{\ampere\per \meter} and \SI{2.7(2)e5}{\ampere\per \meter} respectively.

\begin{figure}
    \centering
    \includegraphics[width=\linewidth]{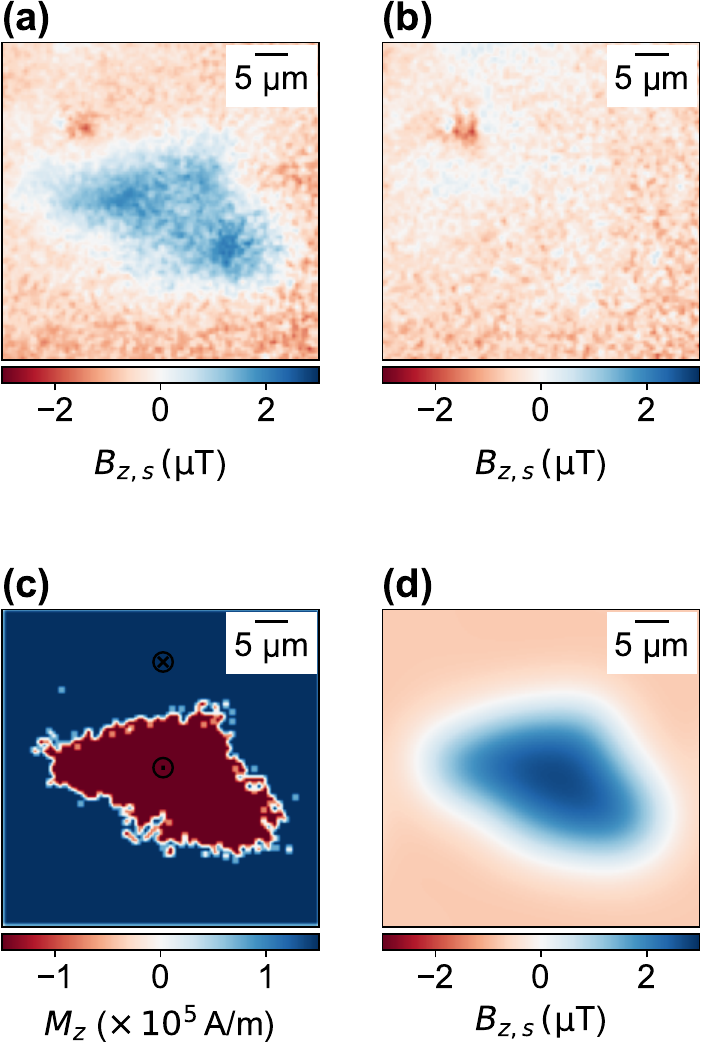}
    \caption{(a). Magnetic field image of the bubble domain captured at \SI{40}{\degreeCelsius} and (b). \SI{50}{\degreeCelsius}. (c). Magnetic domain derived from the bubble domain. (d). Simulated magnetic field corresponding to the bubble domain.}
    \label{fig:domain}
\end{figure}

Although the thin film exhibits predominantly single domain behavior \cite{lin2022chiral,nozaki2020control,tetienne2014nitrogen, mclaughlin2023local,jenkins2019single} under an applied bias magnetic field of \(\SI{0.7}{\milli\tesla}\), a few localized region retains magnetization that is not fully aligned with the applied field. The magnetic field distributions arising from these regions are shown in Fig.\,\ref{fig:domain}(a) and (b) at \(\SI{40}{\degreeCelsius}\) and \(\SI{50}{\degreeCelsius}\), respectively. At \(\SI{50}{\degreeCelsius}\), these localized domain features disappear completely, consistent with the temperature dependent magnetic response observed near the Curie temperature. Such features persist at all other measured temperatures below \(\SI{50}{\degreeCelsius}\).
The corresponding magnetization distribution that can give rise to the observed magnetic field pattern is illustrated in Fig.\,\ref{fig:domain}(c), along with the simulated magnetic field map in Fig.\,\ref{fig:domain}(d), calculated at a sensor stand-off distance of \(\SI{15}{\micro\meter}\). The dots and crosses represent opposite magnetization directions. The magnetization profile resembles that of a 
bubble domain observed in our Pt/Co/Pt thin film. We further characterize the magnetic response in terms of the volume susceptibility, defined as
\begin{equation}
    \chi_{V} = \frac{\Delta{M}}{\Delta{H}}.
\end{equation}
At room temperature, the extracted volume susceptibility is \num{2200}. However, at elevated temperatures, the significantly reduced coercivity makes it difficult to reliably define the volume susceptibility, as the \ce{NV-} measurement is not sensitive near zero fields. Overall, these measurements establish the static magnetic properties of the Pt/Co/Pt thin film, revealing its temperature dependent magnetization, hysteretic switching behavior and localized domain structures. This characterization provides the necessary foundation for investigating the dynamic response of the thin film to time dependent magnetic excitations. 

\begin{figure*}[t]
    \centering
    \includegraphics[width=\linewidth]{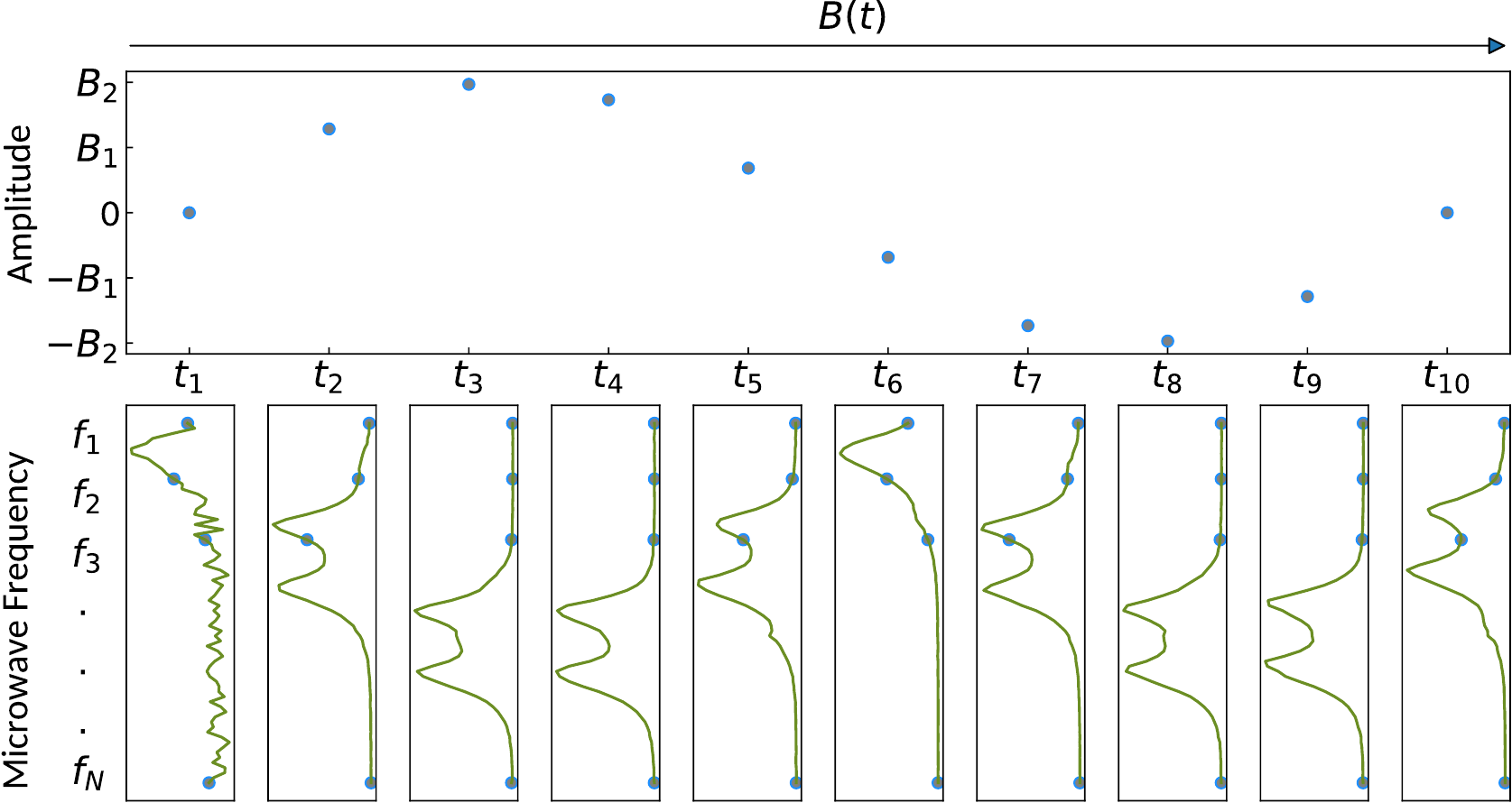}
    \caption{Stroboscopic Method : The magnetic field $B(t)$ is swept by fixing a microwave frequency and collecting the PL response. The same process is repeated for all 
    the microwave frequencies $f_1 \ldots f_N$.}
    \label{fig:schematic}
\end{figure*}

\subsection{Dynamic}

\begin{figure}
    \centering
  \includegraphics[width=1\linewidth]{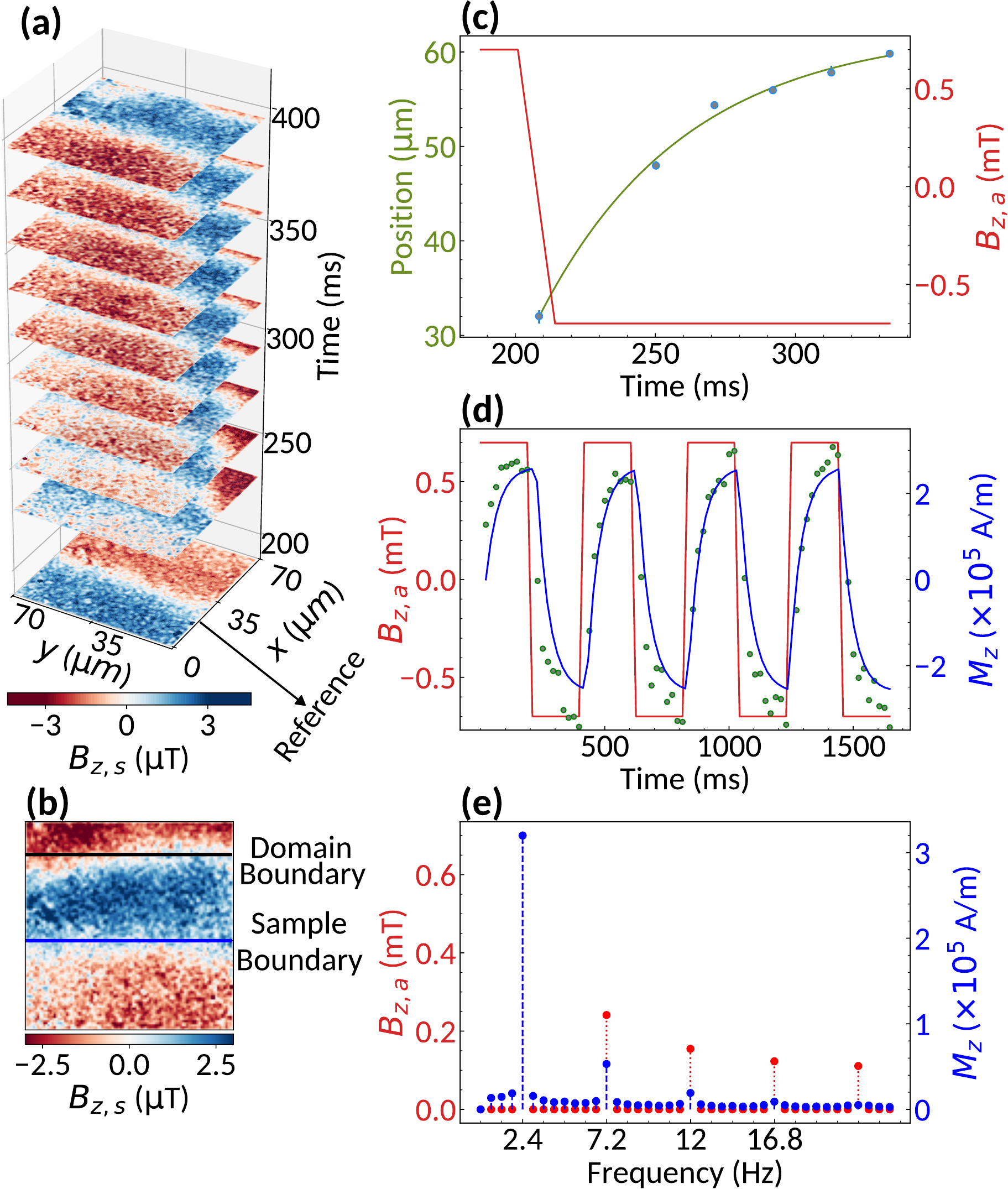}
    \caption{(a). Temporal evolution of magnetic field emanating from
\ce{Pt/Co/Pt} sample as a response to bias magnetic field reversal at \SI{30}{\degreeCelsius}. (b). Magnetic field image 
from the sample at an intermediate stage of reversal. (c). Evolution of domain wall boundary with 
time, fit to an exponential function. (d). Magnetization response to a bipolar square wave 
magnetic field. (e). Fourier transform of the excitation waveform 
and magnetization.}
    \label{fig:time_evolve}
\end{figure}

To investigate the dynamic hysteresis behavior of the thin film, it is necessary to sweep the applied magnetic field between positive and negative saturating fields. Conventionally\,\cite{Parashar2022}, dynamic magnetic imaging using  QDM relies on fixing the microwave frequency at the point of maximum slope in the ODMR spectrum and monitoring the resulting changes in PL intensity. Although this approach enables the detection of small magnetic field (\SI{}{\micro \tesla}) variations around a large bias static magnetic field (\SI{}{\milli \tesla}), its dynamic range is limited by the linewidth $\Gamma$ of the ODMR resonance. If the total magnetic field  changes by more than $\Gamma/\gamma_e$, the change in the
PL becomes nonlinear with change in the magnetic field of the sample, limiting the accuracy of magnetic field measurements. For typical linewidth in the order of \SI{1}{\mega \hertz}, the linear regime in PL change is only around tens of \SI{}{\micro \tesla}
(\SI{35}{\micro \tesla}). In our experiment the field required to saturate the thin film varies from positive \SI{0.7}{\milli \tesla} to negative \SI{0.7}{\milli \tesla}, corresponding to a field variation that significantly exceeds the linear response range with $\Gamma$=\SI{1.6}{\mega \hertz}.
Previously\,\cite{dasika2023mapping}, to overcome
the limited dynamic range of the \ce{NV-} center based magnetic imaging,
a co-sensor was used to track the large changing bias magnetic field, while
the \ce{NV-} sensor was used to track small changes in the magnetic response 
of micron sized micromagnetic sample. However, use of such a scheme is not applicable
in this case because at lower bias fields, the spectra 
from different \ce{NV-} orientations start merging into one another, changing 
the linewidth and amplitude in comparison to higher bias magnetic fields. These spectral changes complicate the calibration and interpretation of the magnetic response, thereby limiting the applicability of the co-sensor approach as well for tracking the complete magnetization reversal process.

To overcome the limited dynamic range, we introduce a stroboscopic measurement technique\,\cite{nasuno2011stroboscopic, li2012stroboscopic} for data acquisition as illustrated schematically in Fig.\,\ref{fig:schematic}.
Unlike conventional ODMR based imaging, which involves fixing the bias magnetic field and sweeping the microwave frequency, a process that typically takes several minutes and limits the ability to capture rapidly evolving magnetic phenomena, we instead, fix the microwave frequency and sweep the magnetic field in synchronization with the desired excitation waveform of our interest at the required frame rate. To observe the dynamics of thin film, we need to vary the bias magnetic field of the excitation waveform faster. This process is repeated for different microwave frequencies to cover the required dynamic range. The data is then organized to obtain PL response as a function of microwave frequency for each bias magnetic field value in the waveform. By fitting a sum of two Lorentzian functions for each
external bias magnetic field, a magnetic field map as shown in Fig.\,\ref{fig:dc}(a),
is obtained. We performed a repeatability check by comparing the magnetic response obtained after averaging over five consecutive cycles,
the response confirms that the magnetization reversal happens by nucleation from the same sites repeatably under
several reversal cycles (see Supplementary Information, Fig. S3). This is in general true as the nucleation 
starts at morphological defects within the thin film\,\cite{10.1063/5.0282414}. However, the domain
dynamics might vary with external field and temperature.

Figure.\,\ref{fig:time_evolve}(a) shows the temporal evolution of the magnetic field generated 
from the thin film in response to a bias square wave magnetic field, illustrating magnetization 
reversal when the applied magnetic field is switched from 
\SI{0.7}{\milli \tesla} to \SI{-0.7}{\milli \tesla}. The base magnetic field map 
acquired at \SI{0.7}{\milli\tesla}, serves as the reference from whereon the bias magnetic 
field is reversed to \SI{-0.7}{\milli\tesla}. The magnetization reversal in this case is a gradual process rather than an instantaneous transition. The magnetic field images are acquired at time interval of 
\SI{20.85}{\milli\second} while maintaining the thin film at
\SI{30}{\degreeCelsius}, the camera binning was maintained at 100 bins and exposure time was set to \SI{1}{\milli \second}. As the bias magnetic field is reversed, the thin film initially nucleates into a multi domain state, while the domains aligned towards the bias magnetic field progressively grows in size across the film. This process continues until the magnetization is completely reversed after \SI{208.5}{\milli\second} from the reference, eventually leaving the sample in a single domain state with its magnetization oriented opposite to its initial direction. The sample boundary and the domain wall boundary are marked clearly in Fig.\,\ref{fig:time_evolve}(b) at an intermediate stage (see Supplementary Information, Fig. S8). Since the thin film is mounted such that the edge of the film should come in the centre of the field of view, so below the thin film boundary it's just bare sensor. Fig.\,\ref{fig:time_evolve}(c) captures the temporal progression of the domain wall in the field of view. The temporal evolution of the domain wall position is phenomenologically given by\,\cite{zhang2021ac,cullity2011introduction}
\begin{equation}
m \frac{d^2x}{dt^2} + \beta \frac{dx}{dt} + \alpha x = 2M_SH,
\label{eq:domain_wall_motion}
\end{equation}
where, $x$ is the position of the domain wall, $m$ is the mass per unit area
of the domain wall, $\beta$ is the viscous damping factor, $\alpha$ is the 
restoring potential, $M_s$ is the saturation magnetization and $H$ is the applied magnetic field. By ignoring the first term\,\cite{cullity2011introduction} in 
Eq.\,\eqref{eq:domain_wall_motion}, which leads to high frequency oscillations, 
we see that the domain wall exhibits an exponentially decaying response to a step change in bias magnetic field with a time constant, $\tau_C = \beta / \alpha$. By fitting the data to an exponentially decaying function, we 
obtain a relaxation time constant of \SI{51}{\milli \second}. The coil current rise time was independently measured to be approximately \SI{3.6}{\milli \second}, which is considerably shorter than the observed relaxation time of \SI{51}{\milli \second}. This indicates 
that the measured relaxation dynamics are primarily governed by the sample response rather than the finite response time of the coil. The physical origin of $\alpha$ is related to crystal imperfections in the 
thin film. As the relaxation time constant is high and the coercivity is low, we assume crystal imperfections in the thin film is very low. Fig.\,\ref{fig:time_evolve}(d), shows the temporal evolution of the sample magnetization, $M_z$ over four consecutive cycles of an applied AC square wave magnetic field ($B_{z,a}$)  having a frequency of \SI{2.4}{\hertz}. The sample response is qualitatively 
identical to the voltage across a capacitor in an RC circuit, driven by a square wave. The result obtained at other temperatures and frequencies are provided in the Supplementary Information (see Fig. S2-S7). As we can drive the thin film across both positive and negative saturation,
where the sample is highly nonlinear, our methodology can also elucidate the nonlinear AC 
susceptibility of the thin film given by
\begin{equation}
M = M_\circ + 	\chi_1 h + \chi_2 h^2 + \chi_3 h^3 + \ldots,
\end{equation}
where, $ M_\circ$ is the spontaneous magnetization, $\chi_1$ is the 
linear susceptibility and $\chi_2, \chi_3, \ldots$ are the higher order
susceptibility coefficients, which can be extracted from the higher order harmonics. The harmonics present in the input and output waveform is shown in Fig.\,\ref{fig:time_evolve}(e). However, for square wave drive the finite relaxation time of the thin film introduces a frequency dependent attenuation of these components, effectively rendering the system as a low pass filter that suppresses higher order harmonics.

\begin{figure}[t]
    \centering
    \includegraphics[width=0.7\linewidth]{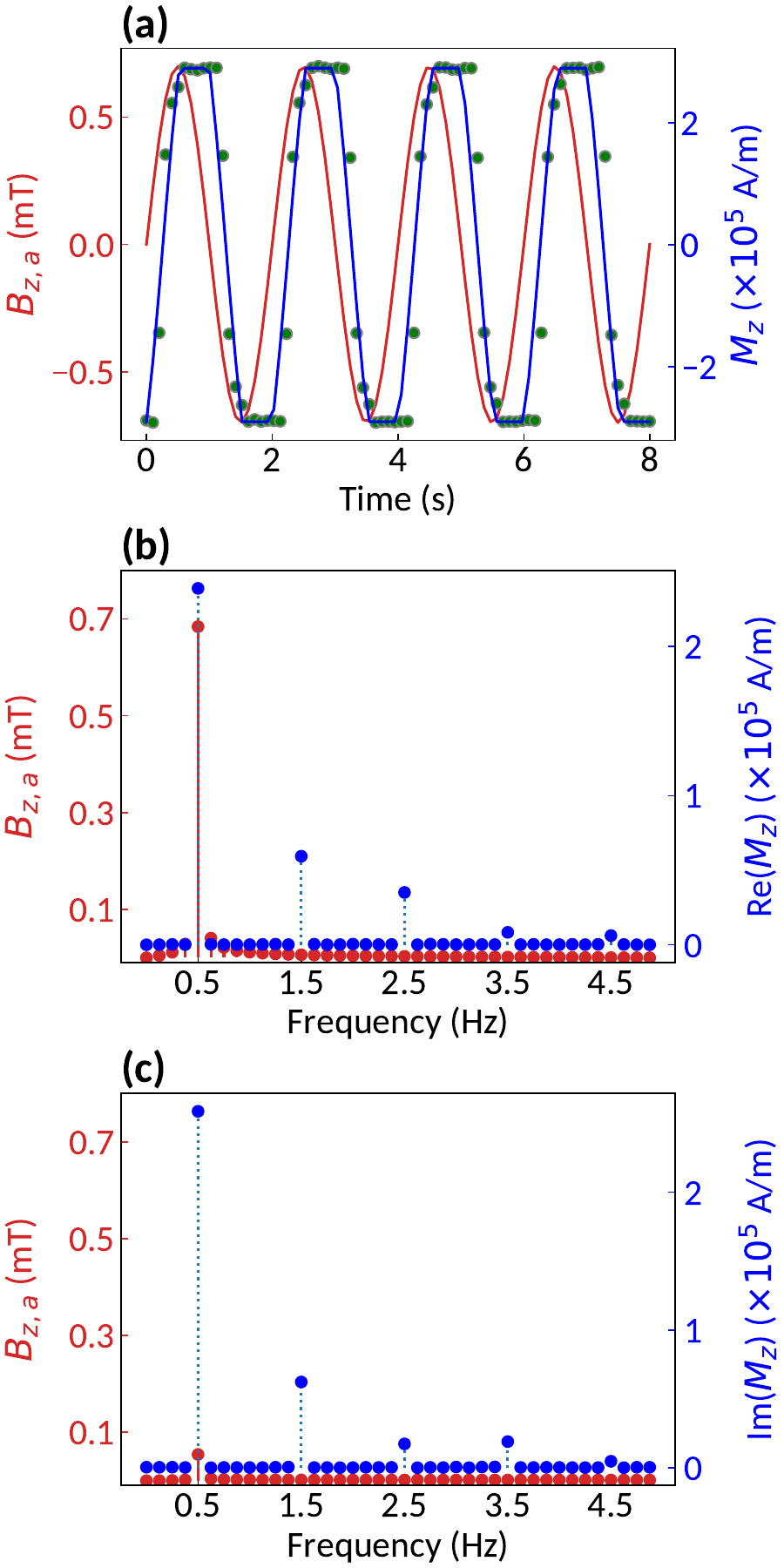}
    \caption{Magnetization response of the thin film 
    due to a sinusoidal bias magnetic field at \SI{30}
    {\degreeCelsius} having a 
    frequency of \SI{0.5}{\hertz} fit to a clipped sinusoidal 
    waveform. (b). Real and 
    (c). imaginary part of the Fourier transform of the excitation waveform and the magnetization response.}
    \label{fig:Fourier}
\end{figure}

Since the square wave drive inherently contains higher order 
harmonics, the resulting magnetization response also comprises of 
contributions from multiple harmonics. To isolate and characterize the higher order harmonics generated by the nonlinear response of the sample, we use a sinusoidal bias magnetic field containing only the fundamental frequency. In Fig.\,\ref{fig:Fourier} (a), we show the magnetization response, $M_z$ of the sample 
at \SI{0.5}{\hertz}, in response to the applied AC sinusoidal magnetic field ($B_{z,a}$) at \SI{30}{\degreeCelsius}, respectively. Despite the sample showing a square hysteresis plot
at \SI{30}{\degreeCelsius} in Fig.\,\ref{fig:dc}(e), the dynamic magnetization response differs significantly at this temperature showing that the magnetization evolves smoothly, following an approximately clipped sinusoidal waveform with a noticeable phase delay of $\phi = 47^\circ$, indicative of a delayed magnetic response (see Supplementary Information, Fig. S5). The phase delay leads to an energy dissipation within the material, given by
\begin{equation}
    E = \mu_\circ \oint H dM = \pi \mu_\circ  M_\circ H_\circ \sin \phi,
\end{equation}

where, $\mu_\circ$ is the permeability of free space, $H_\circ$ is the amplitude of applied magnetic field, $M_\circ$ is the sample magnetization and $\phi$ is the phase delay. The estimated energy dissipated per cycle is E = \SI{5.36e2}{\joule\per\cubic\meter}. In Fig.\,\ref{fig:Fourier}(b),
the real part of the Fourier transform of the bias magnetic field and sample magnetization response is shown. 
The Fourier spectrum contains higher order harmonics due to the clipping in the magnetization response caused by the saturation. Due to the symmetry of the hysteresis curve, the spectrum only contains odd
harmonics. Similarly in Fig.\,\ref{fig:Fourier}(c), the imaginary part of the spectrum is shown, which arises from the delayed response of the magnetization. From these figures, we extract the real first and third order susceptibility coefficients, {$\chi_1^{'}=111.3$} and $\chi_3^{'}= 0.0014~m^2/A^2$ and their imaginary counterparts {$\chi_1^{''}=116.7$} and $\chi_3^{''}=0.0015~m^2/A^2$. The ratio of the cubic and linear terms at the peak field concludes that the sample has strong nonlinear response.  The real or in-phase susceptibility represents the amount of magnetic energy that is stored and recovered during each cycle, corresponding to the reversible magnetization whereas the out-of-phase susceptibility represents the energy absorbed by the material and converted into heat. It reveals energy losses occurring within the system. The ability to simultaneously resolve the phase lag between the applied magnetic field and the resulting magnetization, along with the higher order harmonic components of the response, establishes the potential of this technique as a powerful tool for quantitatively probing nonlinear magnetic dynamics, energy dissipation, and the underlying relaxation processes in complex magnetic materials.

\section{Conclusion}

In conclusion, we demonstrate the capability of \ce{NV^-} based quantum diamond microscope, combined with a stroboscopic measurement technique, to investigate the nonlinear AC response of a Pt/Co/Pt thin film. This approach overcomes the limited dynamic range of conventional optically detected magnetic resonance based imaging, enabling the observation of magnetization reversal under large amplitude bias magnetic field excitations. Our method enables imaging magnetization dynamics on the scales of ms while retaining broad dynamic range and high spatial resolution, establishing it as a promising approach for spatially and temporally resolving magnetization dynamics. The temporal evolution of the magnetic response reveal a gradual magnetization reversal process involving domain nucleation and growth, ultimately leading to a complete magnetization reversal. By tracking the domain wall motion, we extract a relaxation time constant, providing insight into the dynamic response of the thin film. Furthermore, by applying a sinusoidal magnetic field, we detect higher order harmonics in the magnetic response, demonstrating the nonlinear character of the system under AC excitation. The measurements reveal a pronounced phase delay between the applied field and the magnetization response, reflecting the dissipative nature of the system. The observed phase lag together with the magnetization reversal and higher order harmonic content of the response, highlights the potential of this technique for investigating nonlinear magnetic dynamics. Such studies could provide valuable insights into magnetic switching, domain evolution, and dynamical magnetic phase transitions with potential industrial applications in the development of magnetic technologies like random access memories, sensors, etc.

\subsection*{Acknowledgment and Author Contributions}
M.V and D.S conceived and designed the experiment in consultation with K.S. M.V and D.S performed the experiment, while PS. A.K provided
the Pt/Co/Pt thin film sample.  M.V and D.S analyzed the data.  P.S.M aided in extensive data acquisition. M.V wrote the article in discussion with all authors. K.S supervised all aspects of the project.

\subsubsection*{Funding information}
K.S. acknowledges funding from DST National Quantum Mission, and AOARD grant number FA2386-23-1-4012 and FA2386-24-1-4053.

\subsubsection*{Competing Interests}
M.V and D.S declare no competing interests. K.S declares that she is also the project director for the Indian National Quantum Mission's quantum sensing and metrology hub - QMet Tech Foundation.

\subsubsection*{Data availability}
The data supporting the findings of this study are available from the corresponding author upon reasonable request.


%

\end{document}